\documentclass[prb,twocolumn,amsmath,amssymb,aps]{revtex4-1}
\usepackage{times}
\usepackage{graphicx}
\usepackage{bm}
\usepackage{natbib}
\usepackage{textcomp}
\begin{document}
\title{Spin-polarization in the vicinity of quantum point contact with spin-orbit interaction}

\author{Sunwoo Kim}
\author{Yoshiaki Hashimoto}%
\author{Taketomo Nakamura}
\email{taketomo@issp.u-tokyo.ac.jp}
\author{Shingo Katsumoto}%

\affiliation{Institute for Solid State Physics, The University of Tokyo, 5-1-5 Kashiwanoha, Kashiwa, Chiba 277-8581, Japan}

\date{\today}
\begin{abstract}
We have developed a novel technique for detection of spin polarization with a quantum dot weakly coupled to the objective device. The disturbance to the object in this technique is very small since the detection is performed through sampling of single electrons in the object with very slow rate. We have applied the method to a quantum point contact (QPC) under a spin-orbit interaction. A high degree of spin polarization in the vicinity of the QPC was detected when the conductance stayed on a plateau at a half of the unit conductance quantum ($G_{\rm q}/2\equiv e^2/h$), and also on another plateau at $2e^2/h$. On the half-quantum plateau, the degree of polarization $P$ decreased with the bias source-drain voltage of the QPC while $P$ increased on the single-quantum plateau, manifesting that different mechanisms of polarization were working on these plateaus. Very long spin relaxation times in the detector quantum dot probably due to dynamical nuclear spin polarization were observed. Anomalous decrease of $P$ around zero-bias was observed at a Kondo-like resonance peak.
\end{abstract}

\pacs{72.25.Dc, 71.70.Ej, 75.76.+j, 73.63.Kv, 73.63.Nm, 85.35.-p}
\maketitle
\section{Introduction}
Creation and detection of spin polarization in quantum structures of non-magnetic semiconductors are key-techniques in semiconductor spintronics.\cite{awschalom2007challenges} The use of spin-orbit interaction (SOI) is a candidate for the creation of spin current and a number of combinations with quantum structures such as a bent quantum wire,\cite{PhysRevB.72.115321} a quantum point contact (QPC),\cite{doi:10.1143/JPSJ.74.1934} or an Aharonov-Bohm (AB) interferometer\cite{PhysRevB.84.035323} have been proposed though in the authors' view, there has been no report of sound experimental evidence of these devices.

A conductance plateau at a half of quantum conductance $G_{\rm q}$ ($G_{\rm q}\equiv 2e^2/h$, $e$ being the elementary charge, $h$ the Planck constant) in a quantum point contact (QPC) with an SOI (here we call it the ``0.5 anomaly'') was found in experiments~\cite{debray2009all} and considered as the sign of perfectly spin-polarized current. The possible spin polarization is now attracting the attention of researchers for the use as a spin source. However there is no evidence that the current flowing at the 0.5 anomaly is spin-polarized, other than the transport throughout the samples including the shot noise measurement.\cite{kohda2012spin}

A theoretical explanation of the 0.5 anomaly was given by Wan {\it et al.}.\cite{PhysRevB.80.155440} They showed the possibility that the Coulomb interaction and the Rashba-type SOI (RSOI) form a spin-dependent potential, which gives different threshold gate voltages to the spin subbands. The spin-dependence is reversed for the electrons with the opposite momentum, hence this results in a persistent spin-current through the QPCs. Their calculation also shows a kind of spin-standing wave in the vicinity of QPC, which means the appearance of some non-uniform spin accumulation. Another explanation was given by Kohda {\it et al. }, who considered non-uniformity in the effective magnetic field of RSOI due to some gradient in the confinement potential of the QPC.\cite{kohda2012spin} The field gradient gives a direction-dependent force on electron spins, just like the Stern-Gerlach experiment.

Spin polarization due to the RSOI was also predicted on the ordinary conductance plateau at $G_{\rm q}$.\cite{doi:10.1143/JPSJ.74.1934} The polarization in this case is caused by non-equilibrium occupation of quantum levels split by the RSOI. The prediction, however, has not been examined in experiment, mainly because the polarization in this mechanism occurs after electrons pass through the narrowest part of the QPC and no anomaly appears in the conductance. In addition, the small number of polarized electrons are rapidly diluted when they flow out into broader electrodes though the degree of polarization in the vicinity of the QPC should be very high.

According to the fundamental law of symmetry, such a spontaneous spin polarization does not occur in infinite uniform systems, in other words, systems with the spatial translational symmetry, without any mechanism to break the time-reversal symmetry such as external magnetic field or ferromagnetic interaction between magnetic moments. While the former introduces the symmetry breaking directly into single-electron Hamiltonian, the latter brings in spontaneous symmetry breaking through a many-body effect. (Even with the ferromagnetic interaction, it is proved that there is no long range ferromagnetic order in one-dimensional systems, though this issue is out of the present scope.) In finite systems, on the other hand, local spin polarization naturally appears. A quantum dot with an odd number of electrons or in a high-spin state is a representative example. Spontaneous polarization due to the breaking of uniformity, or translational symmetry is also predicted and has been claimed to be the origin of so called the 0.7 anomaly in QPCs with very small SOI.

In this paper we propose a novel method to detect spin polarization with very small disturbance to the objective systems. It utilizes a quantum dot coupled to a side-edge of the system. The dot samples the electrons from the edge with very slow rate, the lower bound of which is determined by the spin-relaxation time in the quantum dot. We have applied the method to QPCs with a moderate strength of spin-orbit coupling and found that high spin polarization rate is obtained not only on the plateau with the conductance $G_{\rm q}/2$ (0.5 plateau) but also on the one with $G_{\rm q}$ (1.0 plateau). The bias dependencies of the polarization on these plateaus were opposite to each other, manifesting the different mechanisms for the polarization. The spin-relaxation time estimated from the duration-controlled experiment is very long probably due to dynamic nuclear polarization when the polarization rate of electrons is high. We also observe Kondo-like enhancement of spin scattering rate around the zero-bias.

\section{Experiment}
\subsection{Principle of polarization detection}
\begin{figure}[b]
\includegraphics[width=\linewidth]{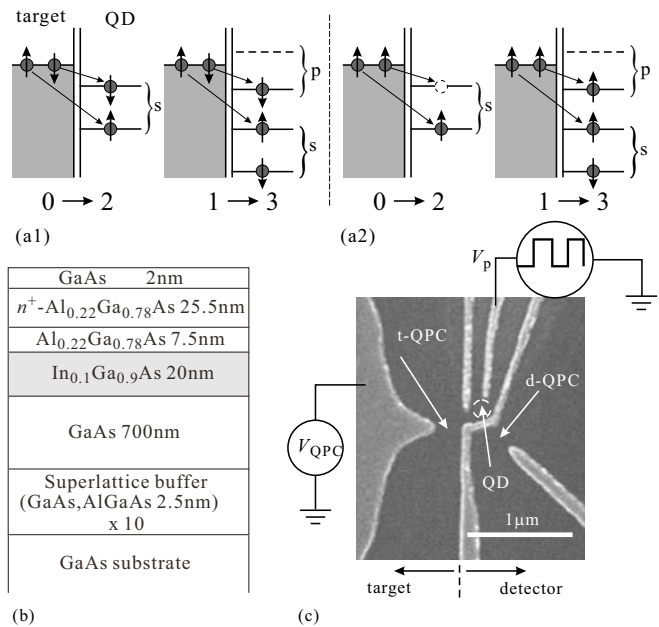}
\caption{\label{fig_exp}
(a1) and (a2) illustrate the principle for detection of spin-polarization by the use of two-electron tunneling processes. (a1) is for a spin-unpolarized target (objective). The Pauli principle does not affect the tunneling rate.
(a2) is for 100 \% ($P=1$) spin polarization. In 0$\rightarrow$2 transition, the Pauli principle blocks the second electron to tunnel. In 1$\rightarrow$3 transition, the second electron is allowed to tunnel when the initially occupying electron has down-spin. ``s'' and ``p'' denote the first orbital state and the second one respectively.
(b) Cross sectional view of the layered structure.
(c) Scanning electron micrograph of a sample showing the gate configuration. Whiter regions are Au/Ti Schottky gates.
}
\end{figure}

The principle of the polarization detection is schematically illustrated in Fig.\ref{fig_exp}(a1) and (a2). The method is based on the measurement of two-electron tunneling rate from an objective into a QD, which is prepared in known electronic states. For rough sketch of the principle, we assume that the total electron concentration in the objective $n_{\rm t}\equiv n_\uparrow+n_\downarrow$ ($n_\sigma,\;\sigma=\uparrow,\;\downarrow$, is the electron concentration with spin $\sigma$) is fixed and that the tunnel coupling between the objective and the QD is common for all the spin and the orbital states. Here we write the averaged time for a tunneling event of a single electron as $\tau_{\rm p}$.

We first consider the case in which the initial state of the QD has no electron (the process 0$\rightarrow$2). When the spin polarization
\begin{equation}
P\equiv (n_\uparrow-n_\downarrow)/(n_\uparrow+n_\downarrow)
\label{eq_polarization_definition}
\end{equation}
in the target device (objective) is equal to 1, {\it i.e.}, the objective has only spin-up electrons, the lowest energy level in the QD denoted as ``s'' in Fig.\ref{fig_exp}(a1) and (a2) can accommodate only one spin-up electron due to the Pauli exclusion principle until the spin flip occurs in the QD. Hence if we take the averaged spin flip time $\tau_{\rm f}$ as infinite, the time-averaged charge $q_{\rm av}$ in the QD over a period $\tau_0$ should be $e(\tau_0-\tau_{\rm p})/\tau_0$ for $\tau_0\gg \tau_{\rm p}$. In the other extreme of $P=0$, $n_\uparrow=n_\downarrow=n_{\rm t}/2$ and $q_{\rm av}$ should be $2e(\tau_0-2\tau_{\rm p})/\tau_0$. When $\tau_0\gg \tau_{\rm p}$, naturally $q_{\rm av}$ is twice of that in the case $P=1$. The two extrema show that $q_{\rm av}$ can work as a measure of $P$. 

In experiments, the number of electrons and $q_{\rm av}$ in QDs can be measured {\it e.g.}, with remote sensing. However, single measurement of $q_{\rm av}$ does not give the value of $P$ directly and some supporting data such as the value of $\tau_{\rm p}$ are required. There are various ways to solve it and here we consider to utilize the tunneling process of 1$\rightarrow$3. For $P=1$ and an initial state of the QD with an up-spin electron, the Pauli principle also blocks the tunneling to the s-state while it does not block the one to the second state (p-state). When the initially occupying electron has down-spin, the tunneling to s-state is allowed. We assume the initial up/down spin states appear with even probability and obtain $q_{\rm av}$ in this case as $3e(\tau_0-\tau_{\rm p})/2\tau_0$. Hence taking the ratio of $q_{\rm av}$ for these two processes, one can obtain the value of $P$.

In more realistic models, the analysis should be more complicated. Within the relaxation time approximation, however, simple analytic formulas can be obtained with solving some rate equations as shown in Appendix.

\subsection{Sample configuration}
A two-dimensional electron system (2DES) in a pseudomorphic ${\rm In_{0.1}Ga_{0.9}As}$ quantum well was grown on a GaAs (001) substrate with ordinary molecular beam epitaxy. Figure \ref{fig_exp}(b) shows the layered structure, in which two GaAs-(In,Ga)As interfaces are placed in the electric field produced by the modulation-doped layer. The RSOI in such 2DES structures is known to be strong due to the asymmetry in the two interfaces.\cite{raey,PhysRevB.68.035315} The carrier concentration and the Hall mobility are $9.8\times 10^{11}~{\rm cm^{-2}}$ and $7.4\times 10^{4}~{\rm cm^2/Vs}$ respectively at 4.2~K. QPCs and a QD were defined by Au/Ti split gates fabricated with electron-beam lithography. Figure \ref{fig_exp}(c) shows a scanning electron micrograph of the metallic gates (whiter regions) with captions being overlapped. The constriction shown in the right side worked as the QPC for remote charge detection (d-QPC) while that in the left side worked as the QPC which showed the 0.5 anomaly (t-QPC). Between these two one-dimensional channels, a QD, which worked as the detector of polarization, was placed and the connection to the left channel existed just next to the narrowest point (slightly upper side in the figure). An advantage in such side-coupled QD structure is the access to $N_{\rm D}=0$ state without killing tunneling conductance to the electrode (in the present case, t-QPC).\cite{doi:10.1143/JPSJ.76.084706}

We fabricated three samples (sample A, B, C) with the same gate configuration. Low-temperature measurements were preformed on the three and essentially the same results were obtained for the spin-polarization on the 0.5 plateaus. Measurements of the spin-polarization for various t-QPC conductances and bias conditions were carried out on one of the three (sample C). The Kondo-like conductance peak was also observed in this sample.

\subsection{Measurement}
The specimens were cooled down to 100~mK in a dilution fridge. A care was taken for the cooling time to be longer than 12 hours from room temperature to 4.2~K. During the cooling process, the split gates were biased at $+0.2$~V for getting leakage-free Schottky gate characteristics with smaller hysteresis in the response of conductance.\cite{PhysRevB.72.115331}

In order to avoid the cross-coupling between the plunger gate and the other terminals, and also to go up to microwave frequencies, the lines to the gates of QD and d-QPC were through coaxial cables. The electrostatic potential of the detector QD was driven by a square wave superposed on the gate voltage. Tunneling of electrons from t-QPC to the QD is detected as the modulation of conductance in d-QPC (remote charge detection). To obtain the charge sensing signal, we adopted
time averaging synchronized to a square wave on the QD gate voltage with lock-in technique.\cite{/content/aip/journal/apl/84/23/10.1063/1.1757023} The remote charge detection is based on the electrostatic sensitivity of d-QPC to the electric field formed by the electrons in the QD. However the gate voltage of QD also modulates the G-V characteristics in t-QPC effectively as shifts in the gate voltage of t-QPC. Thus the modulation was compensated with a software control of the gate voltage in t-QPC for keeping charge sensitivity constant against the gate voltage in QD as far as possible and the residual modulation was normalized with the sensitivity measurement prior to the charge sensing measurement. The signal can thus be transformed to the time-averaged charge $q_{\rm av}$ (with reference to the initial state) with the knowledge of system parameters.

\section{Results and Discussion}
\begin{figure}[b]
\includegraphics[width=0.9\linewidth]{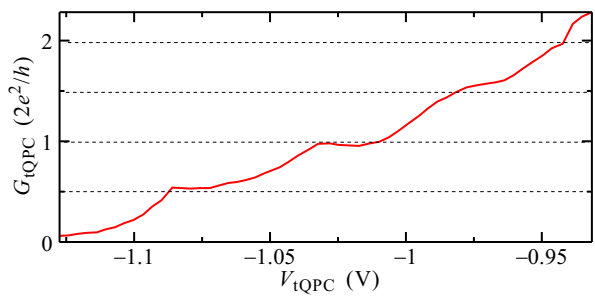}
\caption{\label{fig_tqpc}
Typical gate voltage ($V_{\rm tQPC}$) dependence of conductance through t-QPC ($G_{\rm tQPC}$).
Conductance quantization in units of half conductance quantum ($G_{\rm q}/2=e^2/h$) is observed up to $2G_{\rm q}$.
}
\label{fig_gate_voltage_response}
\end{figure}
Figure \ref{fig_gate_voltage_response} shows a typical gate-voltage dependence of the conductance (G-V characteristics) in t-QPC $G_{\rm tQPC}$. Clear plateau structures appeared around $G_{\rm q}$ and also around $G_{\rm q}/2$ reproducing preceding studies.\cite{debray2009all,kohda2012spin}

\begin{figure}
\includegraphics[width=\linewidth]{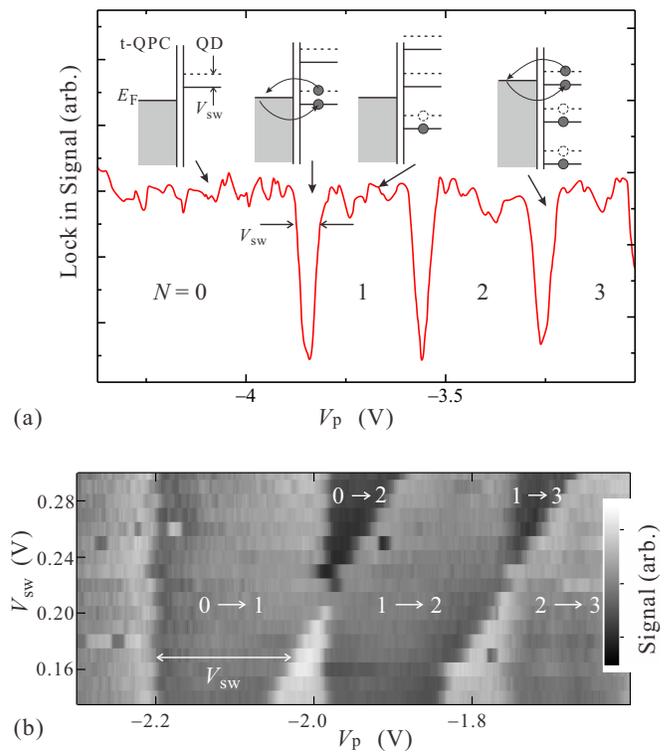}
\caption{\label{fig_signal1}
(a) Typical charge detection signal as a function of the dot plunger gate voltage. The insets are chemical potential diagrams for t-QPC and detector QD. Dip signal appears when $E_{\rm F}$ in t-QPC is within the square wave swing of QD discrete levels hence the dip width is the same as the swing width $V_{\rm sw}$.
(b) Gray scale plot of the signal depth versus the plane of $V_{\rm p}$ and $V_{\rm sw}$. White letters and symbols indicate the corresponding tunneling process with the changes in the electron number. The data are for a different sample from those in (a).
}
\label{fig_charge_detection_signal1}
\end{figure}

Figure \ref{fig_charge_detection_signal1}(a) shows an example of the charge detection signal as a function of the QD gate voltage $V_{\rm QD}$. The signal appears as dips around the points where, as illustrated in Fig.\ref{fig_charge_detection_signal1}(a), the energy (chemical potential) levels in the QD matches with the Fermi energy in t-QPC. As also illustrated in Fig.\ref{fig_charge_detection_signal1}(a), the width of the signal in $V_{\rm QD}$ is the same as the amplitude of the square wave $V_{\rm sw}$. The series of dips ended at $V_{\rm QD}=-1.2$~V, which indicates $N_{\rm D}$ was zero for the lower $V_{\rm QD}$. And we can thus assign the regions in $V_{\rm QD}$ between the dips to $N_{\rm D}$. With increasing $V_{\rm sw}$, the dip width widened with flat bottoms. Hence the signal formed wedge-like regions in the plane of $V_{\rm QD}$-$V_{\rm sw}$ as shown in Fig.\ref{fig_charge_detection_signal1}(b), where the signal is plotted in a gray-scale. The data here were obtained in the open condition of t-QPC, that is, the electrons in the source were not spin-polarized.

If there was some change in the tunneling parameters it should appear as a change in the signal level.\cite{/content/aip/journal/apl/84/23/10.1063/1.1757023} The flatness over the bottoms of the signal regions thus means there was no significant change in the tunneling parameters within the section of $V_{\rm QD}$ in Fig.\ref{fig_charge_detection_signal1}(b).
On the other hand in the overlapped regions of neighboring wedges, clear deepening of the signal due to double-electron tunneling process was observed. We thus can assign single or double electron tunneling process to every wedge-shaped region in Fig.\ref{fig_charge_detection_signal1}(b) as noted with white characters and symbols. Here we represent the tunneling process with the change in $N_{\rm D}$, such as 0$\rightarrow$1 for the process, in which $N_{\rm D}$ changes from 0 to 1. Note that the tunneling rate was strongly reduced and the average time for the tunneling was elongated to 50~{\textmu}s, which corresponds to 3 aA and the net (DC) current was zero.

\begin{figure}
\includegraphics[width=\linewidth]{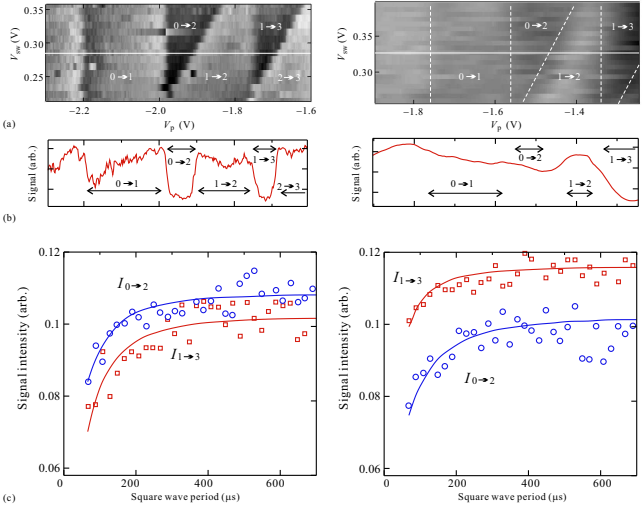}
\caption{\label{fig_spinpolsig}
(a) Gray scale plots of the signal depth for open (left) and 0.5 plateau (right) conditions of t-QPC on the plane $V_{\rm sw}$ versus $V_{\rm p}$. White broken lines in the right panel are just for eyes because the contrast is comparatively weak. The shifts in the abscissae and in the ordinates are due to the electrostatic coupling between the QD and the Schottky gate of t-QPC.
(b) Signal profiles along the horizontal white lines in the gray scale plots in (a). 
(c) Signal intensities at transitions 0$\rightarrow$2 and 1$\rightarrow$3 as a function of period in the square wave driving the QD.
}
\label{fig_comparison_polarization}
\end{figure}

In Fig.\ref{fig_comparison_polarization}, we compare the signal levels for t-QPC opened widely (left panels) and for t-QPC conductance equals $G_{\rm q}/2$ (right panels). The data in the left panels (a) and (b) are very similar to those in Fig.\ref{fig_charge_detection_signal1}(b) besides the shifts due to the difference in the sample characteristics. The signal depths for the processes 0$\rightarrow$2 and 1$\rightarrow$3 are the same within the noise level manifesting that the spin-polarization in wide-opened QPC is almost zero. When t-QPC was pinched for the conductance to decrease and to be quantized at $G_{\rm q}/2$, the pattern changed as in the right panels in Fig.\ref{fig_comparison_polarization}(a), (b), in which an apparent shallowing of the signal for 0$\rightarrow$2 double electron tunneling region was observed. From the qualitative discussion in the previous section, this should come from finite spin-polarization in t-QPC.
Since the electric current through t-QPC was kept to 0 while the measurement, the polarization was not due to the non-equilibrium electric current induced by the boundary conditions (the chemical potentials of the connected reservoirs). Though the SOI does not break the time-reversal symmetry in the total Hamiltonian, it may generate magnetic dipoles resulting in local spin-polarization, which probably has been detected in the present experiment.

To make the analysis more quantitative, we measured the signal intensities for these two regions ($I_{0\rightarrow2}$, $I_{1\rightarrow3}$) as a funtion of the square wave integraion time $\tau_{\rm sw}$ as shown in Fig.\ref{fig_comparison_polarization}(c). When t-QPC was open (the left panel), $I_{0\rightarrow2}$, $I_{1\rightarrow3}$ were at the same level but a small difference due to that in the tunneling rate. On the other hand for 0.5 plateau (the right panel), $I_{1\rightarrow3}$ was ovbiously larger than $I_{0\rightarrow2}$ and the saturation $\tau_{\rm sw}$ was faster for $I_{1\rightarrow3}$ reflecting the spin-polarization.

Applying a simple phenomenological rate equation analysis given in Appendix A, we can fit the data of $I_{0\rightarrow2}(t_{\rm sw})$ $I_{1\rightarrow3}(t_{\rm sw})$ with $P$, $\tau_{\rm sf}$, $\tau_{\rm t}$ as fitting parameters, which can thus be obtained through the analysis. Though the fitting is successful shown as the solid curves, there are some distributions in the data and we need to pay attention to the errors in the fitted values.
We can also obtain the longitudinal spin relaxation time $\tau_{\rm sf}$ in the dot as 200 $-$100$+$250~{\textmu}s. Note that this is so called $T_1$ in the terms of magnetic resonance\cite{Slichter199603}. In weak magnetic fields, that is with finite Zeeman energy, extremely long $T_1$'s reaching ms region have been reported for electron spins in few-electron quantum dots\cite{fujisawa2002allowed,johnson2005triplet,PhysRevLett.91.196802,PhysRevLett.98.126601,doi:10.1143/JPSJ.75.054702}. On the other hand, rapid shrinkage of the relaxation time due to fluctuating nuclear spins has generally been observed with decreasing the energy separation between corresponding spin states to zero. In the present case, the relaxation time of {\textmu}s order is expected referring to preceding results because no external magnetic field is applied and the energy difference is zero. We need to consider, however, that electrons captured by the QD are highly polarized, and dynamic nuclear polarization (DNP)\cite{PhysRevLett.100.067601} should take place inside the QD through the Overhauser effect.\cite{PhysRevLett.99.096804} It is not surprising then, that $\tau_{\rm sf}$ is much longer than {\textmu}s order because fluctuation of nuclear spins in the QD is strongly suppressed by the DNP.

From the fitting we obtained $P$ as 0.7 $\pm$ 0.1 for the right panel in Fig.\ref{fig_comparison_polarization}(c). Note that the measured polarization is for the electrons in the vicinity of sampling gate to the QD and naturally lower than 1 even if the outgoing electrons from t-QPC is perfectly polarized.

\begin{figure}[b]
\includegraphics[width=\linewidth,clip]{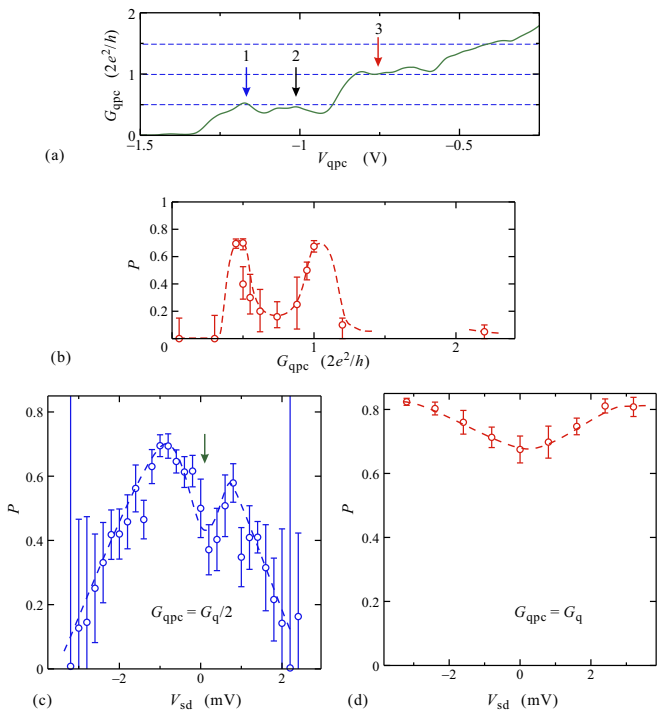}
\caption{(a) Conductance of a t-QPC as a function of the gate voltage. The following data of spin-poalrization has been taken for this sample. Vertical arrows 1 and 3 indicate the positions where the data in (c) and (d) have been taken respectively.
(b) Zero-bias spin polarization as a function of t-QPC conductance. Data $1.5G_{\rm q}\sim2.0G_{\rm q}$ are missing due to experimental limitation. Doubling of the data points at $G_{\rm qpc}=G_{\rm q}/2$ are due to the difference in the results at arrow 1 and 2 in (a), in the former of which we have the Kondo-like conductance enhancement.
(c) Spin-polarization in t-QPC on 0.5 conductance plateau as a function of the source-drain bias voltage ($V_{\rm sd}$). The vertical arrow indicates the zero-bias, where a characteristic decrease in $P$ is observed.
(d) Results of the same measurement as (c) on 1.0 conductance plateau.
\label{fig_sdbiasdep}
}
\end{figure}

High spin polarization $P$ around $G_{\rm qpc}\sim G_{\rm q}/2$ at zero-bias has now been established. We then proceed to measurement of $P$ under other conditions of t-QPC. The conductance of t-QPC $G_{\rm qpc}$ in the sample for this purpose is shown in Fig.\ref{fig_sdbiasdep}(a) as a function of the gate voltage. Quantization at $G_{\rm qpc}\sim G_{\rm q}/2$ is again observed though a peak appears at the gate voltage indicated by the arrow numbered as 1. Figure \ref{fig_sdbiasdep}(a) displays measured $P$ as a function of $G_{\rm qpc}$, which shows an unexpected second peak around $G_{\rm qpc}\sim G_{\rm q}$. By some unknown reason, the measurement was unstable in the section from $1.5G_{\rm q}$ to $2G_{\rm q}$ unfortunately and we could not get reliable data. The doubling of data around $G_{\rm qpc}\sim G_{\rm q}/2$ is due to the difference in $V_{\rm qpc}$ as explained shortly. Note that the sampling of electrons to the QD is through a QPC with conductance much lower than $G_{\rm q}/2$. Hence the sampling itself does not have any function of spin-polarization, which fact is consistent with the present measurement method.

An advantage of side-QD probe is tunability in the position of energy window as reported for the detection of double quasi-Fermi levels in a non-equilibrium quantum wire.\cite{:/content/aip/journal/apl/93/11/10.1063/1.2987424} The present measurement can therefore be applied to the case of finite source-drain bias voltage $V_{\rm sd}$ of t-QPC with tuning the window to local chemical potentials. To investigate the polarization mechanism, we thus measure the variation of $P$ versus $V_{\rm sd}$ on 0.5 and 1.0 conductance plateaus. In this measurement, we tuned the center of sensitivity in our detector QD to the electrode of the opposite side. If the source-drain separation of quasi-Fermi levels is larger than the resolution, we can thus selectively detect the polarization of, {\it e.g.}, electrons passing through the QPC though in the present case the double tunneling energy window is set to be about 2~meV for sufficient signal to noise ratio and clear decomposition was not obtained.

Figure \ref{fig_sdbiasdep}(c) and (d) show the results for 0.5 and 1.0 plateaus respectively exhibiting quite surprising differences in the response to $V_{\rm sd}$. While the polarization decreases with bias voltage and almost vanishes around $V_{\rm sd}\sim 2$~mV on 0.5 plateau, it increases with $V_{\rm sd}$ up to 2~mV on 1.0 plateau. The difference manifests that the spin polarization has been developed through different mechanisms on these plateaus. The data in Fig.\ref{fig_sdbiasdep}(c) is for the gate voltage indicated with arrow 1 in Fig.\ref{fig_sdbiasdep}(a). The small dip in $P$ around zero-bias indicated by the perpendicular arrow in Fig.\ref{fig_sdbiasdep}(c) can be observed only around this gate voltage and disappears for the gate position indicated by arrow 2 in Fig.\ref{fig_sdbiasdep}(a). The zero bias spin-polarization $P$ for this is hence around 0.7 resulting in the data doubling of $P$ at $G_{\rm qpc}\sim G_{\rm q}/2$ in Fig.\ref{fig_sdbiasdep}(b).

\begin{figure}
\includegraphics[width=\linewidth]{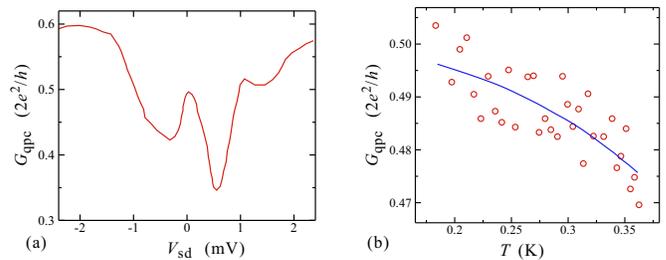}
\caption{\label{fig_kondo}
(a) Bias voltage dependence of the QPC conductance on a 0.5 plateau.
(b) Temperature dependence of the peak conductance shown in (a).
}
\label{fig_kondo_like_peak}
\end{figure}

Hence possible theoretical models for the polarization at 0.5 plateau should be based on spin filtering action, in which a spin-dependent potential at t-QPC blocks either the transport of electrons with up or down spin. In one of such proposals, the transverse confinement potential of QPC works as magnetic field gradient to traversing electron spins (magnetic moments) through the RSOI and causes Stern-Gerlach type force, which works as a spin-dependent potential.\cite{PhysRevB.72.041308,kohda2012spin} In another proposal, RSOI in combination with the electron-electron Coulomb interaction forms spin-dependent effective potential.\cite{PhysRevB.80.155440} They are in qualitative agreement with the present experimental results in that finite $V_{\rm sd}$ would overcome such spin-dependent potential resulting in decrease of $P$. In Fig.\ref{fig_kondo}(a), we show the differential conductance of t-QPC as a function of $V_{\rm sd}$. As indicated by the arrow in Fig.\ref{fig_kondo}(a), at zero-bias a conductance peak structure appears corresponding to the dip in $P$ shown in Fig.\ref{fig_sdbiasdep}(c). Besides the zero-bias peak, $G_{\rm qpc}$ increases with $|V_{\rm sd}|$ due to surmounting the barrier of QPC.\cite{PhysRevB.62.10950} The increment in $G_{\rm qpc}$ is in the range from 0.5 to 2~mV in $V_{\rm sd}$ in accordance with the decrement in $P$ supporting the above inference. We cannot go to further judgment though, which theory is more plausible, the both being consistent with the present result.

The zero-bias conductance peak shown in Fig.\ref{fig_kondo_like_peak}(a) is reminiscent of the Kondo peak often observed around 0.7 anomaly in QPCs with small SOI.\cite{PhysRevLett.88.226805} The peak is observed only at the gate position indicated as 1 in Fig.\ref{fig_sdbiasdep}(a). The width of the peak is roughly 100~{\textmu}V, which corresponds to about 1~K in temperature. The peak conductance increases with lowering the temperature as shown in Fig.\ref{fig_kondo_like_peak}(b). It is possible to fit the well-known Kondo temperature dependence\cite{0953-8984-6-13-013} to the data with the Kondo temperature $T_{\rm K}\sim 1$~K as indicated by the blue curve. The fridge condition prevented us from further confirmation of the Kondo effect and we thus call the conductance peak as ''Kondo-like" peak henceforth. The spin polarization in Fig.\ref{fig_sdbiasdep}(c) shows a dip structure around zero-bias corresponding to the conductance peak indicating that the Kondo resonance enhances the spin scattering and disturbs the polarization of traversing electrons.

At small biases, the polarization recovers the value as high as 0.7. The result means the Kondo-like effect has narrower bias window than the spin filter effect manifesting that there is no direct relation between the origins of these two effects but the competition through the spin scattering. There is some asymmetry in Fig.\ref{fig_kondo_like_peak}(a) for the zero-bias. This is presumably because the detector energy window is tuned to the electrode in the opposite side. A small bias voltage gives imperfect separation of two quasi-Fermi levels and would enhance the polarization detected when it is not as large as to overcome the spin-dependent potential.

On the other hand at 1.0 plateau, the spin filtering mechanisms cannot be applied because there is no significant conductance shift from the original quantized value. Instead some spin rotation process should be considered to elucidate the spin polarization. To our knowledge there is only one such theoretical proposal,\cite{doi:10.1143/JPSJ.74.1934} in which the SOI and the transverse confinement potential cause level anti-crossings between dispersion branches with different spins resulting in spin rotation of traversing electrons. The direction dependent spin rotation generates spin polarizations on both ends of the QPC with inverse polar directions, one of which would be detected in our measurement because the electron exchange point is slightly off from the narrowest point of the QPC. From our measurement, then, a magnetic quadrupole would be formed around the QPC. The polarization by the spin rotation mechanism should not change very much as far as the quasi-Fermi level exists above the level anti-crossing point. The small and almost symmetric enhancement of polarization with the finite bias then again would come from separation in the quasi-Fermi levels and tuning of the energy window to the electrode across the QPC. As the separation increases, the spin polarization detector gradually focuses on the electrons which have passed the top of the QPC potential.

\section{Summary}
In summary, we have measured the spin polarization of electrons in the vicinity of quantum point contacts with strong spin-orbit interaction by using side-coupled quantum dot spin detectors. The polarization is as high as 0.7 on the conductance plateaus of 0.5 and 1.0 times conductance quantum 2$e^2/h$. The polarization decreases with finite biasing on the former while it increases on the latter indicating that a spin filtering effect is working in the former and a spin rotation is working in the latter.

This work was supported by Grant-in-Aid for Scientific Research on Innovative Area, ``Nano Spin Conversion Science" (Grant No.26103003), also by Grant No.25247051 and by Special Coordination Funds for Promoting Science and Technology.

\section*{Appendix}
Let $N_{\rm t}\equiv N_\uparrow+N_\downarrow$ be the total number of electrons energetically available and spatially close enough to tunnel into the QD. Then from \eqref{eq_polarization_definition}, $N_\uparrow=(1+P)N_{\rm t}/2$, $N_\downarrow=(1-P)N_{\rm t}/2$.

\subsection*{0$\leftrightarrow$2 process}
In the bias region of 2$\leftrightarrow$0 and in the charge up process of the QD, the tunneling channels for up and down spins can be treated independently due to the Pauli principle and the rate equations for $q_\sigma(t)$ (averaged charge for spin $\sigma$ electrons on the QD) are written as
\begin{equation}
\left\{
\begin{aligned}
\delta q_\uparrow(t)&=[(e-q_\uparrow(t))N_\uparrow\gamma_0+(q_\downarrow-q_\uparrow)\gamma_f]\delta t,\\
\delta q_\downarrow(t)&=[(e-q_\downarrow(t))N_\downarrow\gamma_0+(q_\uparrow-q_\downarrow)\gamma_f]\delta t
\end{aligned}
\right.
\label{eq_02_charge_rate}
\end{equation}
for the evolutions $\delta q_{\uparrow\downarrow}$ in infinitesimal time $\delta t$, where $\gamma_f$ is the spin relaxation rate
in the QD. For simplicity, the tunneling probability is fixed to $\gamma_0$.
In eq.\eqref{eq_02_charge_rate}, the crossing terms of $q_\uparrow q_\downarrow$ cancel out.
These rate equations are readily solved to give $q(t)=q_\uparrow(t)+q_\downarrow(t)$ as
\begin{equation}
q(t)/e=
2+A^+\exp(\lambda^+t)
+ A^-\exp(\lambda^-t),
\end{equation}
where $A^\pm$ are the integration constants and
\begin{equation}
\lambda^\pm\equiv-(N_t\gamma_0+\gamma_f)\pm\xi,\quad\xi\equiv\sqrt{(PN_t\gamma_0)^2+\gamma_f^2}.
\label{eq_lambda_def}
\end{equation}
The discharging process 2$\rightarrow$0 can be viewed as the charging process of ``hole" and the rate equations are
\begin{equation}
\left\{
\begin{aligned}
\delta q_\uparrow(t)&=[-q_\uparrow(t)N_\uparrow\gamma_0+(q_\downarrow-q_\uparrow)\gamma_f]\delta t,\\
\delta q_\downarrow(t)&=[-q_\downarrow(t)N_\downarrow\gamma_0+(q_\uparrow-q_\downarrow)\gamma_f]\delta t.
\end{aligned}
\right.
\label{eq_02_discharge_rate}
\end{equation}
Eq.\eqref{eq_02_discharge_rate} are essentially the same differential equations as eq.\eqref{eq_02_charge_rate}.

Imposing the boundary condition of the square-wave gate-voltage with period 2$\tau$, duty ration 50\%, the result
\begin{subequations}
\begin{multline}
\frac{q(t)}{e}=
2-\left(\dfrac{\gamma_f}{\xi}+1\right)\dfrac{\exp(\lambda^+t)}{\exp(\lambda^+\tau)+1}\\
+\left(\dfrac{\gamma_f}{\xi}-1\right)\dfrac{\exp(\lambda^-t)}{\exp(\lambda^-\tau)+1}, \quad 0\leq t <\tau,
\end{multline}
\begin{multline}
\phantom{\frac{q(t)}{e}}=\left(\dfrac{\gamma_f}{\xi}+1\right)\dfrac{\exp[\lambda^+(t-\tau)]}{\exp(\lambda^+\tau)+1}\\
-\left(\dfrac{\gamma_f}{\xi}-1\right)\dfrac{\exp[\lambda^-(t-\tau)]}{\exp(\lambda^-\tau)+1}, \quad \tau \leq t <2\tau
\end{multline}
\label{eq_02_semi_final}
\end{subequations}
is obtained.

Because the signal $S$ is obtained from the lock-in technique, we need to perform the integral calculation
\begin{equation}
S\propto \int_0^{2\tau}q(t)\cos\left(2\pi\frac{t}{\tau}\right)dt.
\label{eq_lock_in_signal}
\end{equation}
This can be analytically done\cite{/content/aip/journal/apl/84/23/10.1063/1.1757023} though we avoid to write down a long tedious expression.

\subsection*{1$\leftrightarrow$3 process}
Addition of an energy level makes the calculation in this process much more complicated.
Let the average charge of the upper $p$-orbital as $q_3$. 
For simplicity we use the fact that the energy relaxation from the $p$-state to the lower $s$-state is very fast\cite{zibik2009long}.
Then we write down the rate equations for the charging process on the assumptions: 1) there is a tunneling process with a constant rate to the lower $s$-state through the upper $p$-state; 2) the charge of the upper $p$-state does not increase until the lower $s$-state is full;
as 
\begin{subequations}
\begin{align}
\dfrac{dq_3}{dt} &=\dfrac{q_\uparrow q_\downarrow}{e}(e-q_3)N_t\gamma_3,\label{eq_upper_prob}\\
\dfrac{dq_\uparrow}{dt} & =(e-q_\uparrow)N_\uparrow\gamma_2+(q_\downarrow-q_\uparrow)\gamma_f,\\
\dfrac{dq_\downarrow}{dt} & =(e-q_\downarrow)N_\downarrow\gamma_2+(q_\uparrow-q_\downarrow)\gamma_f.
\end{align}
\label{eq_probs}
\end{subequations}

Even with this approximation, eq.\eqref{eq_upper_prob} contains non-linear terms 
though the other two differential equations in \eqref{eq_probs} can be independently solved.
Hence with substituting the solutions to \eqref{eq_upper_prob} we can obtain the analytical results.
\begin{align}
q_\uparrow&=e+A_{11}\exp(\lambda_3^+t)+A_{12}\exp(\lambda_3^-t),\\
q_\uparrow&=e+A_{21}\exp(\lambda_3^+t)+A_{22}\exp(\lambda_3^-t),
\end{align}
where $A_{ij}$ are constants and $\lambda_3^\pm$ can be obtained from $\lambda^\pm$ in \eqref{eq_lambda_def} replacing $\gamma_0$ with
$\gamma_2$. And the first equation in \eqref{eq_probs} is now written as
\begin{multline}
-\frac{edq_3}{q_3+e}=(e+A_{11}\exp(\lambda_3^+t)+A_{12}\exp(\lambda_3^-t)\\
\times (e+A_{21}\exp(\lambda_3^+t)+A_{22}\exp(\lambda_3^-t))dt.
\end{multline}
The RHS can be readily integrated and LHS is $-\log|1+q_3/e|$ though, again we avoid writing down the tedious solution.

In the discharging process, with reversing the above idea for approximation, the discharging rate from the upper level is a constant and
\begin{equation}
q_3(t)=q_{30}\exp(-\lambda_{3{\rm d}}t).
\end{equation}
Then the other two are
\begin{align}
\dfrac{dq_\uparrow}{dt}&=(1-q_3/e)N_\uparrow\gamma_2q_\uparrow +(q_\downarrow-q_\uparrow)\gamma_f,\\
\dfrac{dq_\downarrow}{dt}&=(1-q_3/e)N_\downarrow\gamma_2q_\downarrow+(q_\uparrow-q_\downarrow)\gamma_f.
\end{align}
Writing down the solutions of the above equations requires special functions.
Boundary condition fitting and the signal calculation on \eqref{eq_lock_in_signal} should be done numerically.

\subsection*{Practical fitting process}
Fitting with the above obtained results are too time consuming. Instead we first adopt empirical formulas 
to obtain the estimation of the tunneling rates.

\noindent For {0$\leftrightarrow$2 process}
\begin{multline}
S\propto\left[ \left\{ \left(1+\frac{\gamma_f}{\xi} \right)\frac{\pi^2}{\lambda^{+2}\tau^2+\pi^2}\left(1+\exp(\tau\lambda^+)\right)-2 \right\}\right.\\
+\left.\left\{ \left(1-\frac{\gamma_f}{\xi} \right)\frac{\pi^2}{\lambda^{-2}\tau^2+\pi^2}\left(1+\exp(\tau\lambda^-)\right)-2\right\} \right].
\end{multline}
For {1$\leftrightarrow$3 process}
\begin{multline}
S\propto\left[ \left\{ \left(1+\frac{\gamma_f}{\xi} \right)\frac{\pi^2}{\lambda_3^{+2}\tau^2+\pi^2}\left(1+\exp(\tau\lambda_3^+)\right)-2 \right\}\right.\\
+\left.\left\{ \left(1-\frac{\gamma_f}{\xi} \right)\frac{\pi^2}{\lambda_3^{-2}\tau^2+\pi^2}\left(1+\exp(\tau\lambda_3^-)\right)-2\right\} \right]\\
+\chi_2\left[ \frac{\pi^2}{\lambda_{3{\rm d}}^2\tau^2+\pi^2}-1 \right],
\label{eq_4_13cds}
\end{multline}
where $\chi_2$ is a fitting parameter determined by the boundary condition.
The obtained values are checked with substitution to the original equations.

\end{document}